# Pulse-shape discrimination in water-based scintillators

Michael J. Ford[a]*, Natalia P. Zaitseva[a], M. Leslie Carman[a], Steven A. Dazeley[a], Adam Bernstein[a], Andrew Glenn[a], Oluwatomi A. Akindele[a]*

[a]*Lawrence Livermore National Laboratory*
*7000 East Avenue, Livermore, CA 94551*

**Abstract**

This work describes a class of liquid scintillators that contain mostly water (>50 wt. % of the entire composition) and can discriminate between interactions induced by neutrons and gamma rays. By balancing the interface interactions between the components of the formulation, these scintillators form emulsions that can be thermodynamically stable. This approach, which considers a quantity known as the hydrophilic-lipophilic difference, requires consideration of the salinity and temperature as well as characterization of the surfactants and oil phase. Emulsions comprised of water and various oils were characterized first. Then, the effect of scintillating dyes in the oil phase was considered, followed by the construction of partial phase diagrams of the emulsions. For transparent oil-in-water emulsions with a single phase, the scintillation light yield and properties of pulse-shape discrimination were measured. The best performing scintillators contained 33 wt. % of a scintillating oil phase and exhibited a light yield that was as high as 18% of the light yield of a commercially available liquid scintillator that does not contain water (EJ-309). These water-based liquid scintillators exhibited a figure of merit of neutron/gamma ray discrimination as high as 1.79 at about 1500 keV$_{ee}$.

---

[*] This is to indicate the corresponding author.
 Email address: akindele1@llnl.gov; ford40@llnl.gov

1. Introduction

In some respects, water is an ideal medium for large detectors (≥ kiloton scale)[1,2]. It has high optical transparency, has minimal impact on the environment, is easy to handle and transport, and is affordable when compared to organic scintillators. Detection of ionizing radiation in pure water relies on Cherenkov radiation[3]. Comparatively, organic scintillators have higher light yield and lower detection threshold. Another advantage of organic scintillators is their ability to discriminate between ionizing radiation (e.g., gamma rays vs. neutrons). Water-based liquid scintillator (WbLS) has been proposed as an intermediate option for large detectors. WbLS could exhibit higher light output and improved sensitivity to ionizing radiation of lower energy when compared to pure water while potentially maintaining much of the optical transparency of pure water[4].

WbLS incorporates organic scintillator into water, relying on a surfactant to form an emulsion (**Figure 1a**). Since WbLS was introduced, researchers have made meaningful progress on scintillator characterisation and development of detector applications of WbLS[4–7]. The ability to separate Cherenkov and scintillation signals has been identified as a key advantage of WbLS[7]. Large detectors (up to 50 kiloton) comprised of WbLS have been considered for potential applications in neutrino and proton decay physics[8]. While a few formulations have been explored[6,9,10], WbLS formulations commonly contain linear alkyl benzene (LAB) as the oil phase and the sulfonated derivative of LAB, linear alkyl benzene sulphonic acids, as the surfactant. Further understanding of the balance between the surfactant, oil phase, and water phase will be crucial for developing WbLS and improving scintillation performance and stability.

As described in previous reports, the phase behaviour of any emulsion is dictated by the ionic strength of the aqueous phase (S), the temperature (T, in ºC), the hydrophobic or hydrophilic nature of the surfactant (Cc), and the lipophilicity of the oil (the effective alkane carbon number, EACN)[11]. For ionic surfactants, the effect of these four parameters can be quantitatively expressed as a dimensionless quantity, the hydrophilic-lipophilic difference (HLD):

$$HLD = Cc - k(EACN) - \alpha(T - 25°C) + \ln S \qquad (1)$$

In this equation, $k$ is generally about 0.15-0.17, depending on the surfactant, and $\alpha$ is 0.01 °C$^{-1}$. The ionic strength, $S$, is a combination of the ionic strength of the salt added and the ionic strength of the surfactant[12]. For a formulation with HLD < 0 and with equal content of oil and water, an oil-in-water emulsion is favoured, and phase separation occurs for any excess volume of oil that does not emulsify. For a formulation with HLD > 0 and with equal content of oil and water, a water-in-oil emulsion is favoured, and any excess volume of water separates. For a formulation with HLD ≈ 0 and with equal content of oil and water, three phases can form, including one where interactions between the surfactant and the oil and water phases are equally favourable, maximizing the oil-water interfacial area[13].

HLD has gained traction as a useful tool for formulating emulsions that are thermodynamically stable (e.g., for oil recovery)[12]. For a set of components that form an emulsion, it's important to know all quantities that influence HLD. These quantities allow a formulator to create stable emulsions where the dispersion phase and continuous phase are predictable (oil-in-water or water-in-oil). Simple experiments can be used to estimate unknown quantities in the HLD equation. For example, if the EACN of the oil is unknown, surfactants with known values of Cc can be mixed to scan across a wide range of Cc and then determine EACN (i.e., a Cc scan).

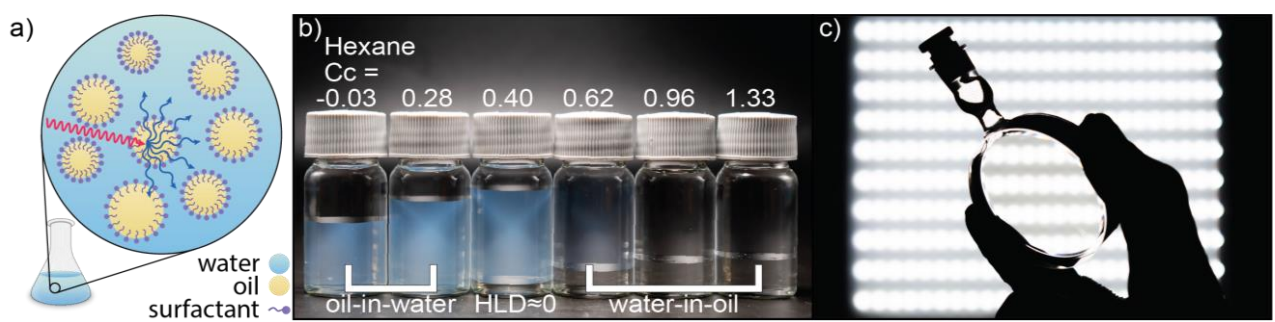

**Figure 1.** a) For WbLS, ionizing radiation (red arrow) is converted to visible light (blue arrows) in the oil phase that is stabilized by surfactants, as shown in this cartoon. b) A scan of Cc can estimate EACN by identifying where HLD ≈ 0, as show

in this photograph. The vials are 28 mm in diameter. c) Transparent emulsions for scintillators can be formulated, as shown in this photograph of a cuvette filled with a WbLS that contains an oil phase comprised of 8 wt. % 2,5-diphenyloxazole and 0.5 wt. % 9,10-diphenylanthracene in LAB. The overall oil content is about 33 wt. % and the overall water content is about 50 wt. %. For scale, the cuvette has a diameter of 50 mm.

To illustrate, emulsions of hexane and water were formulated using different weight ratios of two surfactants. The surfactants sodium dodecylbenzene sulfonate (SDBS, $Cc$ = -0.9) and sodium dioctyl sulfosuccinate (DOSS, $Cc$ = 2.55) provide a wide range of accessible values of $Cc$[14]. The HLD ≈ 0 when the $Cc$ = 0.40 (**Figure 1b**). For $Cc$ = 0.40, Eq. 1 can be set such that HLD = 0. The remaining values needed to solve Eq. 1 to get EACN are known. Using this equation, the EACN is about 6, which agrees with the reported value for hexane[13].

We've attempted to quantify HLD for various oil phases that could be used for WbLSs. In mapping partial phase diagrams of WbLS formulations, we discovered several formulations that form single phases that are transparent or semi-transparent (e.g., **Figure 1c**). The light yield predominately depends on the oil content. Pulse shape discrimination (PSD) is possible for WbLS with an appropriate oil content (about 10 wt. % and greater). These results highlight possible detection capabilities of WbLS and are a promising step toward further development of large detectors using WbLS.

## 2. Materials and methods

All materials were used as received without further purification except m-terphenyl and 9,10-diphenylanthracene, which were purified by recrystallization or washing. Hexane, ethyl benzene, propyl benzene, hexyl benzene, m-terphenyl, 2-butanol, 9,10-diphenylanthracene, sodium dodecylbenzene sulfonate, and sodium dioctyl sulfosuccinate were purchased from Sigma. Toluene, cyclohexane, and sodium chloride were purchased from VWR. 2,5-diphenyloxazole was purchased from Acros Organics. Water that was purified through a Milli-Q water purification system was used. Stock solutions of the water phase were prepared by dissolving and mixing sodium chloride salt (typically 1.5 g/100 mL water) and 2-3 wt. % of 2-butanol. When referencing the water phase, we include the content of sodium chloride and 2-butanol. Typically, all surfactants were weighed into individual vials, and then liquids were added by mass or volume. Stock solutions of scintillators were prepared by dissolving appropriate weight content of scintillating dyes into organic solvents. All contents of the formulations were dissolved by shaking and heating to about 60 °C for a period of one day, followed by further shaking as necessary. Before taking photographs, formulations were allowed to settle and equilibrate over multiple days. Detailed recipes of formulations that were used to determine HLD characteristics of LAB are provided in the Supporting Information as examples (**Table S1**).

All photographs were taken using a Nikon D750 and were globally edited in Adobe Lightroom for colour and exposure corrections. Photographs of fluorescing scintillators were taken while the WbLS was illuminated by a 365 nm UV lamp. To highlight scattering due to emulsion formation, photographs were taken using front lighting and backlighting. These photographs were used to estimate the HLD. Where three phases occurred, this formulation had an HLD value of approximately zero. Other phases were identified as HLD < 0 or HLD > 0 based on whether the characteristic of the surfactant was more hydrophilic or hydrophobic than the characteristics of the surfactant used for HLD = 0. These qualitative characterizations were also used to estimate the phase diagrams of the formulations.

Before scintillation characterization, the WbLS formulations were stirred overnight in an oxygen-free environment to remove dissolved oxygen unless otherwise indicated. WbLS formulations were pipetted into cylindrical cells with quartz windows 50 mm in diameter with a 10 mm path length (Starna Cells). The outer edge and one face of the cells were wrapped and covered with Teflon tape. The exposed face was coupled with optical grease to a Hamamatsu R6231-100-SEL photomultiplier tube (PMT). The PMT interfaced with a 14-bit high resolution CompuScope 14200 waveform digitizer and signals from the PMT were recorded at a sampling rate of 200 MS/s. Ionizing radiation from $^{60}$Co incident upon the WbLS produced scintillation that provided a measurement of light yield of the WbLS. Comparison with $^{137}$Cs was used for confirmation of the energy of ionizing radiation. From histograms of the pulse integral, the Compton edges of $^{137}$Cs and $^{60}$Co of scintillators could be observed. The location of 500 keVee was defined by the pulse integral at 50% of the height of the $^{137}$Cs Compton edge, which could be equated to the Compton edge of $^{60}$Co at 1118 keV using a calibration curve that was generated from various organic scintillators.

For PSD, the WbLS formulations were also exposed to a $^{252}$Cf source shielded behind 5.1 cm of lead to reduce the gamma-ray flux. The waveforms from measurements with $^{252}$Cf were integrated over two separate time integrals to determine the total charge ($Q_{total}$), while the later time integral comprised the charge of the delayed component of the signal ($Q_{tail}$). Since scintillation pulses that result from interactions induced by neutrons are generally longer in time, the charge of the delayed component relative to the total charge is generally greater for scintillation that occurs due to neutrons and small for scintillation that occurs due to gamma-rays. The PSD can be quantified using a Figure of Merit (FoM) that is determined from histograms of the ratio of the charge of the delayed component relative to the total charge, as described in previous reports[15,16]. Briefly, the FoM is:

$$FoM = \frac{\langle n \rangle - \langle \gamma \rangle}{FWHM_n + FWHM_\gamma} \qquad (2)$$

In this equation, $\langle n \rangle$-$\langle \gamma \rangle$ represents the difference between the average value of the neutron and gamma-ray signals, and $FWHM_n + FWHM_\gamma$ represents the sum of the full-width at half of the maximum value of the distributions of the neutron and gamma-ray signals.

## 3. Results and discussion

### 3a. Studies of emulsion formation

We focused on the HLD of alkylated benzenes. These aromatic solvents can participate in energy transfer to facilitate scintillation. Non-aromatic solvents like hexane and cyclohexane do not participate in energy transfer but were also studied for comparison of emulsion formation (**Figure 1b, Figure S1**). For aromatic solvents, we performed Cc scans with SDBS and DOSS on toluene, ethyl benzene, propyl benzene, and hexyl benzene (**Table 1**). These scans involved systematically changing the relative weight content of SDBS and DOSS and estimating where HLD ≈ 0 based on photographs of the emulsions (**Figure 2a** for summary of HLD**, Figure S1** for photographs). For toluene, the surfactant scan was not sufficient to estimate EACN. The Cc at which HLD ≈ 0 was not in the range that we studied. All emulsions that formed were water-in-oil with excess water separating to the bottom. For ethyl benzene, the EACN is around -0.5. As the number of carbons on the alkyl chain increases, the EACN increases as expected (i.e., the oil phase becomes more lipophilic). For propyl benzene, the EACN is around 2, and for hexyl benzene, the EACN is around 6.

**Table 1:** Summary of Cc scans for different solvents. HLD ≈ 0 occurred when a third phase appeared and was used to estimate the EACN (**Figure S1**). These formulations contained 2-3% 2-butanol and 1.5 g/100 mL NaCl in the water phase.

| Solvent | Cc values scanned | Cc where HLD ≈ 0 | EACN |
|---|---|---|---|
| Hexane | -0.03, 0.28, 0.40, 0.62, 0.96, 1.33 | 0.40 | 6 |
| Cyclohexane | -0.62, -0.33, -0.11, 0.06, 0.32, 0.62 | -0.33 | 2 |
| Toluene | -0.85, -0.76, -0.62, -0.54, -0.22, 0.12 | N/A | <-2 |
| Ethyl benzene | -0.85, -0.76, -0.62, -0.54, -0.22, 0.12 | -0.62 | -0.5 |
| Propyl benzene | -0.76, -0.54, -0.22, 0.12, 0.45, 0.82 | -0.22 | 2 |
| Hexyl benzene | -0.76, -0.54, -0.22, 0.12, 0.45, 0.82 | 0.45 | 6 |
| Linear alkyl benzene | 0.62, 0.96, 1.33, 1.72, 2.12, 2.55 | 1.33 | 11 |

Given that linear alkyl benzene (LAB) is common in organic liquid scintillators and is used in previously reported WbLS formulations, its phase behavior in emulsions is important to consider. LAB is commercially available as a mixture of alkylated benzenes. The number of carbons of the alkyl substituents for the LAB used in this report (Alkylate 225, purified by Eljen Technology) is 10-16. The EACN of LAB as estimated by a Cc scan is 11 (**Figure 2a, Figure S2**). This EACN could be compared to the EACNs of ethyl benzene, propyl benzene, and hexyl benzene, which have alkyl substituents containing 2, 3, and 6 carbons respectively. For LAB, the EACN is greater than the EACN for these other alkylated benzenes, consistent with the trend that increasing the number of carbons on the alkyl substituent increases the EACN.

In some formulations (e.g., ethyl benzene or propyl benzene), we observed opaque phases. These are either due to large micelles or complicated phase behaviour (e.g., formation of liquid crystal phases by the surfactants).[14] Smaller micelles scatter blue light. When observed with backlighting, these emulsions with smaller micelles appear transparent (**Figure S1**). To formulate transparent emulsions, small micelles must be formed, which increases the oil/water interface area. One way to decrease micelle size would be to increase surfactant concentration. The molecular structure of the oil phase and surfactant may also play a role in the micelle size. For example, when comparing hexane and cyclohexane at HLD ≈ 0, the middle phase for the hexane emulsion is more transparent than the middle phase for the cyclohexane emulsion at an identical surfactant concentration. We speculate that this difference in transparency may relate to: a) the nature of the surfactant mixture since a formulation containing hexane at HLD = 0 includes more DOSS, which has two flexible hydrocarbon chains that can readily interact with the oil phase, and/or b) the nature of the oil since hexane is an extended alkane that may be more accessible to the surfactant due to flexibility and sterics. Future work should explore these ideas, but they suggest that LAB may be preferable as the primary oil phase for WbLS formulations that contain DOSS and SDBS, given the EACN of LAB and the extended nature of the alkyl substituents.

*3b. Scintillating emulsions*

The emulsions described above cannot be used as WbLSs because they lack scintillation dyes, which may change the phase behavior of the emulsions. For example, the addition of 1 wt. % 2,5-diphenyloxazole (PPO) to LAB resulted in the emergence of a middle phase at Cc = 0.96 that persisted at Cc = 1.33. For comparison, a middle phase for undoped LAB was observed only at Cc = 1.33 (**Figure 2b** for summary of HLD, **Figure S2** for photographs). Thus, the EACN for 1 wt. % PPO in LAB is about 10 (vs. 11 for undoped LAB, **Table 2**). For 5 wt. % PPO in LAB, the EACN further decreased to about 9. For comparison, 5 wt. % m-terphenyl (mTP) in LAB shifted the EACN to 10. PPO contains nitrogen and oxygen, whereas mTP does not contain any heteroatoms, which may influence the EACN. For an oil phase that contains 15 wt. % PPO in LAB, the EACN again decreased to about 6. The introduction of a secondary dye (e.g., 9,10-diphenylanthracene, DPA) at 0.5 wt. % to LAB had minimal effect on the HLD/EACN. To summarize, the Cc at which HLD = 0 decreases as PPO content increases, indicating that PPO requires a more hydrophilic surfactant mixture to balance the emulsion (**Figure 2b**). These results emphasize that formulation of WbLSs must consider all components of each phase since undoped LAB will have different interactions with surfactants than LAB that contains dissolved dyes.

**Table 2:** Summary of Cc scans for different oil phases contained LAB and scintillating dyes. HLD ≈ 0 occurred when a third phase appeared and was used to estimate the EACN (**Figure S2**). These formulations contained 2-3% 2-butanol and 1.5 g/100 mL NaCl in the water phase.

| Solvent | Cc values scanned | Cc where HLD ≈ 0 | EACN |
|---|---|---|---|
| Linear alkyl benzene | 0.62, 0.96, 1.33, 1.72, 2.12, 2.55 | 1.33 | 11 |
| 1 wt. % PPO in LAB | 0.28, 0.62, 0.96, 1.33, 1.72, 2.12 | 1.33 | 10 |
| 5 wt. % PPO in LAB | -0.03, 0.28, 0.62, 0.96, 1.33, 1.72 | 0.96 | 9 |
| 5 wt. % mTP in LAB | -0.03, 0.28, 0.62, 0.96, 1.33, 1.72 | 1.33 | 10 |
| 15 wt. % PPO in LAB | -0.33, 0.28, 0.45, 0.53, 0.62, 0.96 | 0.53 | 6 |
| 15 wt. % PPO + 0.5% DPA in LAB | -0.03, 0.28, 0.62, 0.96, 1.33, 1.72 | 0.62 | 6 |

Thus far, HLD was evaluated to determine EACN, and this evaluation requires the content of the oil phase to equal the content of the water phase. To facilitate scaling of detectors to large sizes, it is desirable to have water comprise the majority of WbLS. Thus, it is also important to construct phase diagrams for compositions of interest and elucidate phase behaviour of the WbLS formulations. We prepared various formulations that contained LAB with 15 wt. % PPO as the oil phase. Phase diagrams were estimated from formulations with different values of Cc (-0.24, 0.35, 0.53, 1.04), which influence the HLD (ca. -0.8, -0.3, 0, 0.5). Phases that were observed include: oil in water with excess oil on top (Type I); water in oil with excess water on bottom (Type II); a microemulsion of oil and water that is sandwiched between excess oil and excess water on top and bottom (Type III, **Figure 2c**).

The prevalence of Type I vs. Type II in the phase diagram depends on the HLD. For HLD < 0, Type I was favourable. As HLD increases, Type II becomes more favourable (**Figure 2d**). This preference highlights an important aspect of using HLD to formulate WbLS: by changing the HLD, the function of the WbLS could change. For example, radioactive species that are water soluble can be analysed in liquid scintillation cocktails that are commercially available. These cocktails contain high contents of aromatic oils and small amount of surfactant to emulsify water-soluble radioactive species. For these cocktails, HLD > 0 may be more favourable to create water in oil emulsions (Type II) that use surfactants most efficiently. The same oil phase can be used for Type II emulsions as for Type I emulsions if the surfactancy balance is accounted for. Future work in this area should consider principles of net average curvature, which can provide a predictive approach for formulating a given emulsion[17].

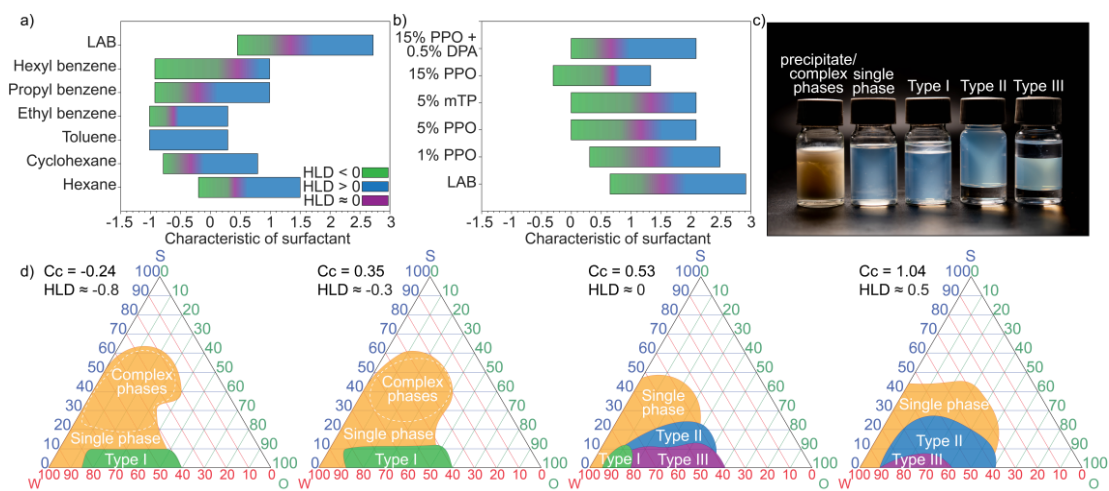

**Figure 2.** a) The HLD of an emulsion is influenced by the EACN, as shown in this plot that summarizes the phase behaviour of emulsions containing different oil phases as the Cc of the surfactant was changed. b) The EACN also changes when dyes are added, as shown in this plot that summarizes the phase behaviour of emulsions containing LAB and various dyes. For a) and b), the HLD was estimated from photographs of the Cc scans (available in the Supporting Information). A gradient colour plot is used to highlight the estimated HLD of the formulations that were studied; HLD greater than and less than 0 should extrapolate as Cc increases or decreases. c) When evaluating the phase diagrams of these emulsions, various phases were observed, as shown in these photographs of emulsions in 28 mm vials. Photographs were taken with front lighting, and blue scattering is apparent. d) Phase diagrams at different values of Cc were constructed. As the Cc increases, a Type II emulsion (water in oil) becomes more favourable.

Some formulations were highly opaque, were highly viscous, contained solid particles, and/or became coagulated formed due to complex phases (e.g., liquid crystal phases)[14] or due to precipitation (**Figure 2d**). These formulations are not suitable for scintillation. Single phases that are transparent are ideal for WbLS. In estimating the phase behaviour of these WbLS formulations, we identified formulations that contained single phases that would be suitable as WbLSs (**Figure 3a,b**). The viscosity and stability of these single phases varied; however, the use of additional 2-butanol in the water phase could help with stability in some cases without influencing the HLD (**Figure S3**). In addition to WbLSs based on LAB with ionic surfactants, we also formulated WbLSs that contained doped propyl benzene with ionic surfactants and doped xylene with a nonionic surfactant (Igepal ® Co-720). These formulations were tested for light output and, in some cases, PSD.

### *3c. Scintillation characterization*

An important feature of WbLS is the light yield produced by the organic liquid scintillator phase. The relative weight content of the oil phase that is comprised of organic liquid scintillator dictates the light yield of a WbLS. As oil content increases, the relative light yield increases. Light yields as high as 18% of the light yield of an organic liquid scintillator that does not contain water (EJ-309) were observed (**Figure 3c**). Consistent with our results, other reports of this previously reported WbLS highlighted higher light yield at higher content of scintillator[10]. In our formulation, the oil phase comprises 33 wt. % of the total scintillator. The surfactant and water phase comprise 17 wt. % and 50 wt. % of the total scintillator respectively. As typical for liquid scintillators, the removal of dissolved oxygen by sparging with nitrogen may improve light yield. Generally, the relative weight content of the water phase inversely affects the light yield when compared to the relative weight content of the oil phase (**Figure 3d**). Notably, the light yield can reach 12% of the light yield of EJ-309 when the water phase comprises > 60 wt. % of the WbLS and can reach 9% of the light yield of EJ-309 when the water phase comprises > 80 wt. % of the WbLS. These values can provide a baseline for future improvements in WbLS formulation. The relative weight content of the surfactant shows no general trend for light yield. Higher content of surfactant generally means that more oil can be dissolved; lower light yield at higher surfactant content indicates that some emulsions may not be using surfactant efficiently (**Figure 3e**).

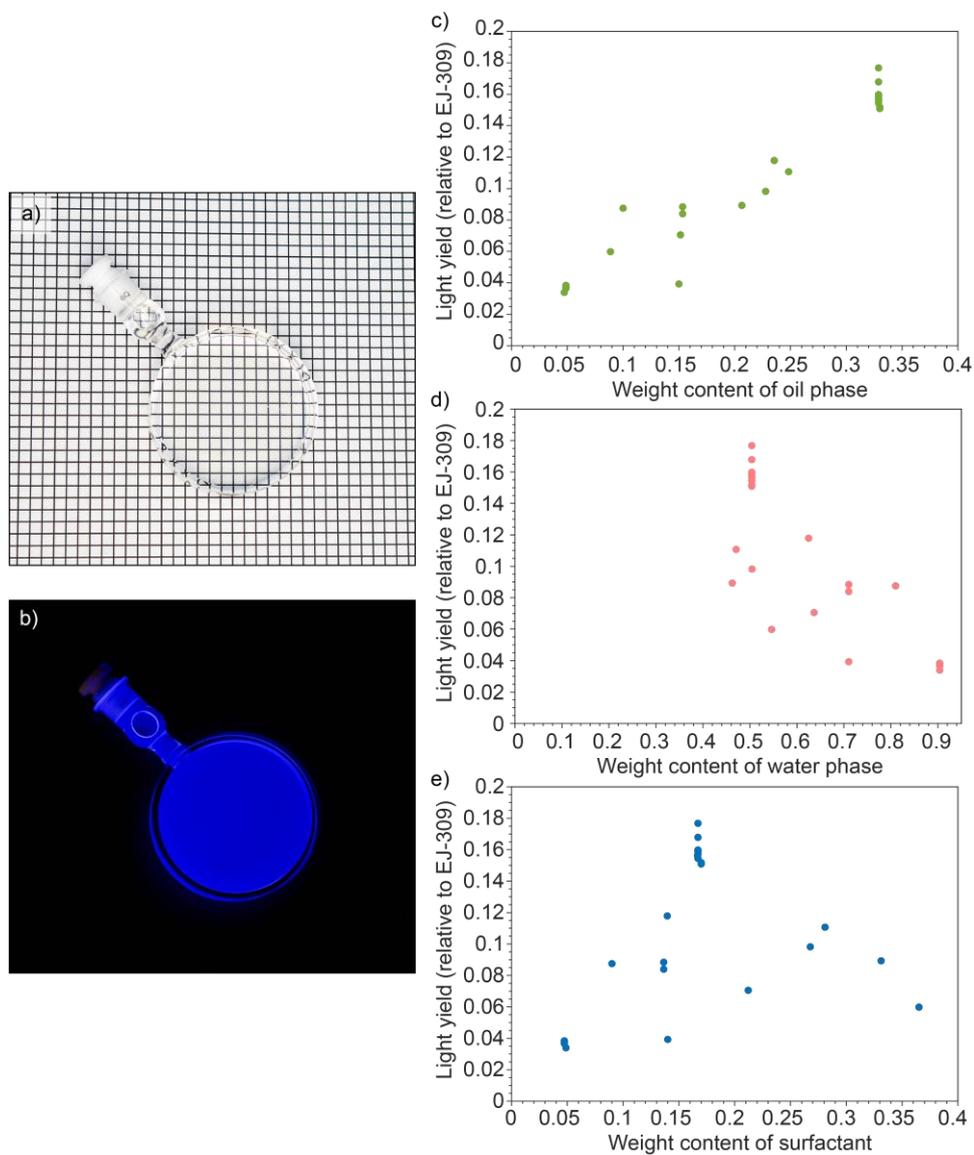

**Figure 3.** a), b) Formulations that formed a single, transparent phase were suitable for scintillation, as shown in these photographs of a cuvette containing a WbLS. The overall oil content is about 33 wt. % and the overall water content is about 50 wt. %. The photograph in b) was taken with the WbLS under a UV lamp. For scale, the cuvette has a diameter of 50 mm. c) The light yield generally increases as the content of the oil phase increases, as shown in this plot. d) The light yield generally decreases as the content of the water phase increases, as shown in this plot. e) There is no general

trend for light yield vs. surfactant content, as shown in this plot. All plots contain multiple data points for a given value on the x-axis (e.g., oil) since the remaining weight content of a WbLS is comprised of two components that can vary relative to one another (e.g., water and surfactant).

In addition to affecting light output, the weight content of the oil phase in WbLS influences the ability to discriminate between gamma rays and neutrons. Discrimination of gamma rays and neutrons is possible by comparing the delayed component of scintillation light vs. the total scintillation light; neutrons generally produce a larger fraction of delayed component than gamma rays. Scintillation from WbLS with higher weight content of oil (e.g., 33 wt. % oil phase) has clear separation of gamma rays (**Figure 4a**). As the weight content of the oil phase decreases, the separation is less clear. For a WbLS with 15 wt. % oil phase, the scintillation signatures of neutrons and gamma become broader, and PSD is apparent but less obvious than for a WbLS with 33 wt. % oil phase (**Figure 4b**). Higher energy particles are necessary to observe PSD at lower weight content of the oil phase. Once the oil content is at 5 wt. %, PSD is no longer apparent (**Figure 4c**).

The FoM for PSD quantifies the effect of the oil content (**Figure 4d**). As the weight content of the oil phase increases from 5 to 15 to 33 wt. %, the FoM increases from about 0.50-0.65 to 0.88-1.20 to 1.23-1.53. Note that the FoM depends on energy and increases moderately as energy increases for a given WbLS. The FoM is further improved when incorporating propyl benzene as the primary organic solvent instead of LAB. For WbLS that incorporates propyl benzene, PSD is apparent (**Figure 4e**). A FoM is as high as 1.79 has been obtained for the gamma-equivalent energy of 1500 keV$_{ee}$. The mechanism for this increase in FoM for PSD will be explored as we continue to develop our formulations.

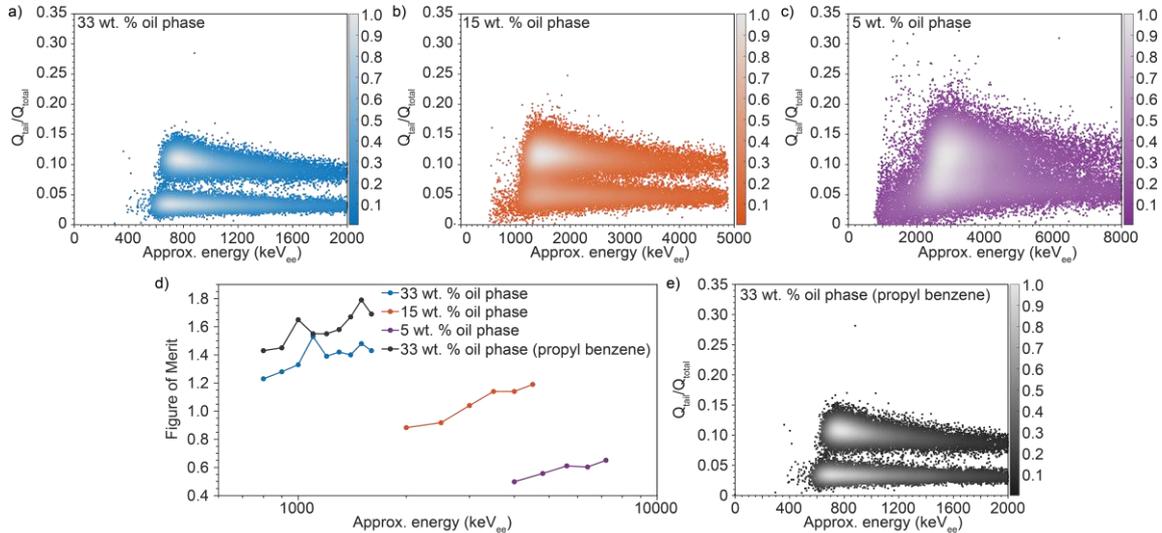

**Figure 4.** The gamma/neutron PSD characteristics can be compared for WbLS formulations that contain a) 33 wt. % oil, b) 15 wt. % oil phase, and c) 5 wt. % oil phase. d) The Figure of Merit depends on energy for a given WbLS and decreases as oil content decreases. e) PSD was observed for a WbLS with an oil phase that contained propyl benzene as the organic solvent. Gradient scale bars represent relative density of data points.

## 4. Conclusion

A key advantage of WbLS is the ability to tune the content of the oil phase to suit the detection needs[10]. Specific applications may rely on properties of pure water (e.g., low cost, directionality of Cherenkov radiation) or properties of organic liquid scintillators (e.g., light yield), and both phases can coexist in WbLS. Notably, WbLS can identify particles based on the signatures of Cherenkov radiation and organic liquid scintillation. Here, we establish an additional functionality that can be achieved in WbLS: pulse-shape discrimination. PSD can help with particle identification and thus reduce cosmogenic fast neutron backgrounds or reduce systematic uncertainties that may arise when using large detectors that rely on WbLS. By mapping partial phase diagrams of various WbLS formulations, we also highlight the importance of formulating using a rational approach like HLD. That is, the HLD can be adjusted as another way to suit detection needs. This approach allows one to formulate and favour oil-in-water emulsions over water-in-oil emulsions or vice versa, depending on the application requirements. With these principles established for WbLS, future work should focus on materials development to further understand scintillation properties and particle discrimination. Formulations could also be formed using a predictive approach like net average curvature[17]. By improving the functionality of WbLS, we hope to facilitate future development of large particle detectors.

## Acknowledgements

This work was performed under the auspices of the U.S. Department of Energy by Lawrence Livermore National Laboratory under Contract DE-AC52-07NA27344, LDRD tracking number 21-FS-019, release number LLNL-JRNL-829820. We thank Prof. Steven Abbott and Prof. Edgar Acosta for useful discussions.

The authors declare no competing interests.

# Supporting Information: Pulse shape discrimination in water-based scintillators

Michael J. Ford[a], Natalia P. Zaitseva[a], M. Leslie Carman[a], Steven A. Dazeley[a], Adam Bernstein[a], Oluwatomi A. Akindele[a*]

[a]*Lawrence Livermore National Laboratory*
*7000 East Avenue, Livermore, CA 94551*

**Table S1.** Detailed recipe of the formulations used to determine hydrophilic-lipophilic difference (HLD) characteristics of LAB. The surfactants sodium dodecylbenzene sulfonate (SDBS, Cc = -0.9) and sodium dioctyl sulfosuccinate (DOSS, Cc = 2.55) were used to determine the effective alkane carbon number (EACN) of LAB.

| SDBS mass (g) | DOSS mass (g) | NaCl mass (g) | 2-butanol mass (g) | water mass (g) | oil mass (g) | CC value | "Type" |
|---|---|---|---|---|---|---|---|
| 0.4 | 0.4 | 0.108 | 0.216 | 6.876 | 8 | 0.62 | I |
| 0.32 | 0.24 | 0.108 | 0.216 | 6.876 | 8 | 0.96 | I |
| 0.24 | 0.32 | 0.108 | 0.216 | 6.876 | 8 | 1.33 | III |
| 0.16 | 0.64 | 0.108 | 0.216 | 6.876 | 8 | 1.72 | II |
| 0.08 | 0.72 | 0.108 | 0.216 | 6.876 | 8 | 2.12 | II |
| 0 | 0.8 | 0.108 | 0.216 | 6.876 | 8 | 2.55 | II |

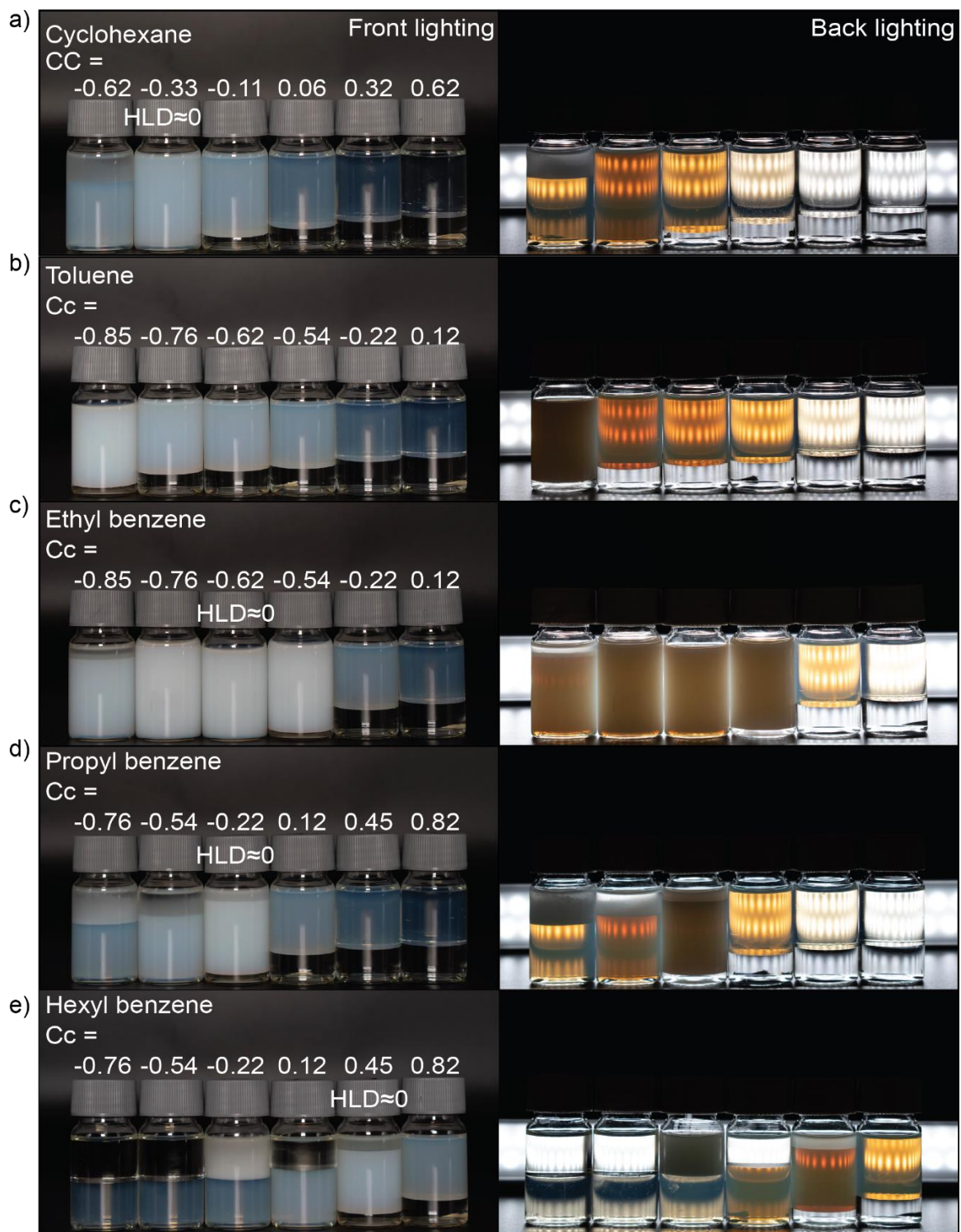

**Figure S1.** By varying the weight content of two surfactants with different Cc, the overall Cc can be changed until the HLD ≈ 0. This Cc scan can estimate EACN of various oils, as show in these photographs taken with front lighting (left) and backlighting (right) for cyclohexane (a), toluene (b), ethyl benzene (c), propyl benzene (d), and hexyl benzene (e). These photographs were referenced when filling Table 1. The vials are 28 mm in diameter.

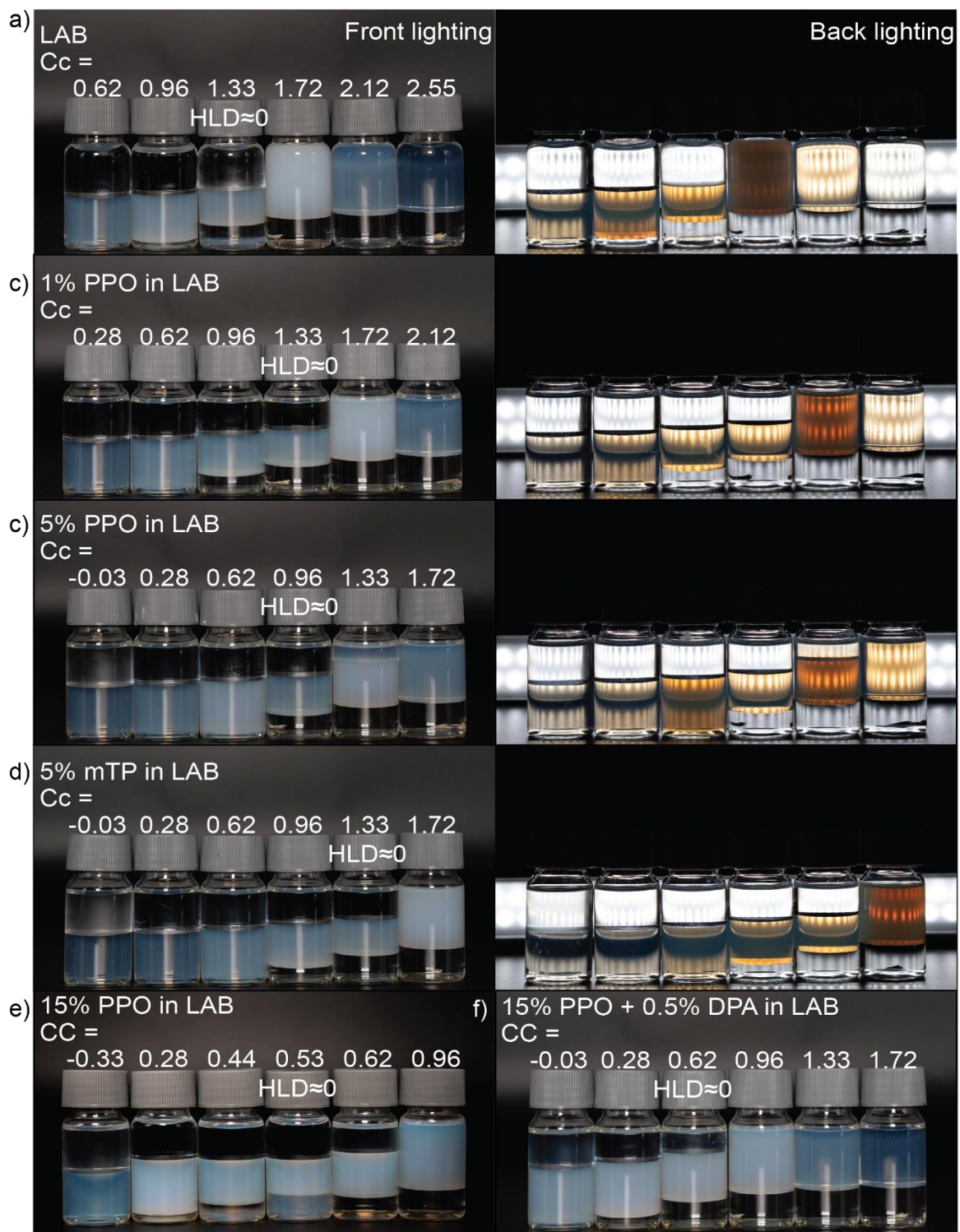

**Figure S2.** By varying the weight content of two surfactants with different Cc, the overall Cc can be changed until the HLD ≈ 0. This Cc scan can estimate EACN of LAB containing various amounts of scintillating dye, as show in these photographs taken with front lighting (left) and backlighting (right) for undoped LAB (a), 1 wt. % PPO in LAB (b), 5 wt. % PPO in LAB (c), 5 wt. % mTP in LAB (d), 15 wt. % PPO in LAB (e), and 15 wt. % PPO along with 0.5 % DPA in LAB (f). These photographs were referenced when filling Table 2. The vials are 28 mm in diameter.

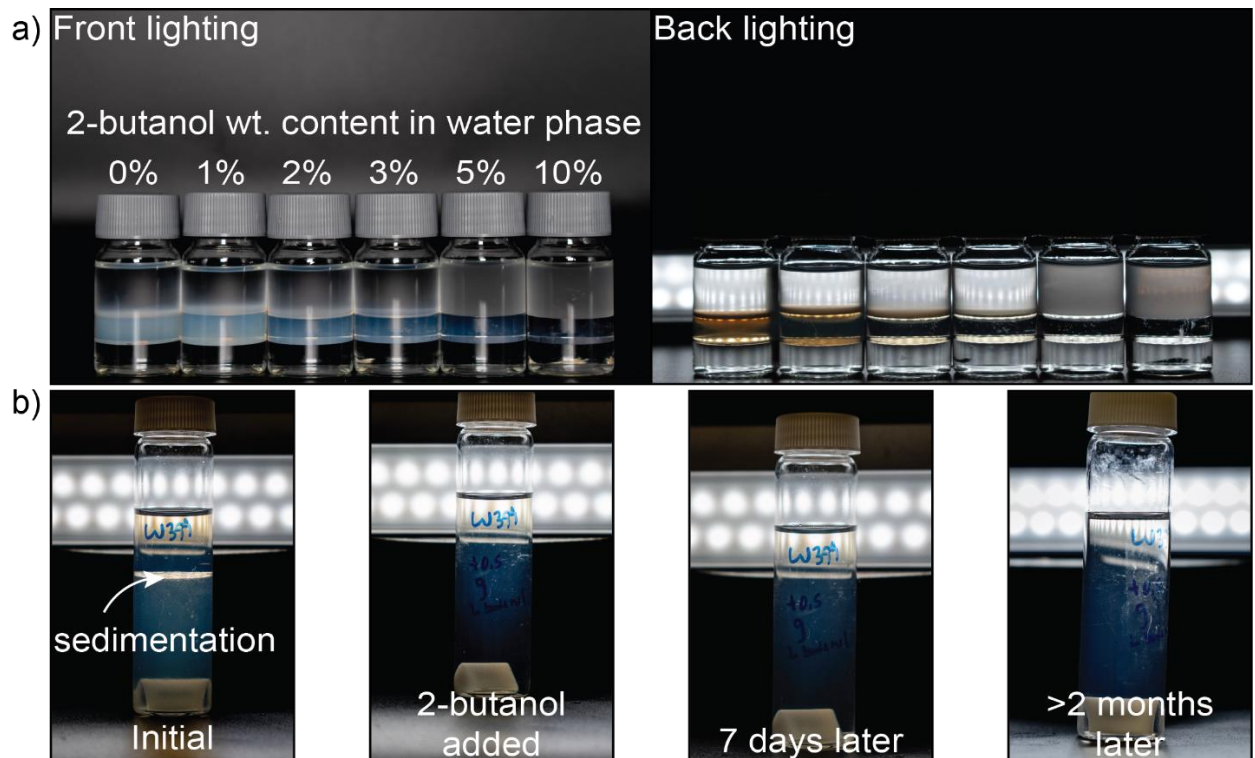

**Figure S3**. a) The addition of up to 10 wt. % 2-butanol does not have a meaningful effect on HLD, as shown in these photographs taken with front lighting (left) and backlighting (right) of scintillation vials that contain a formulation where HLD is close to zero (i.e., three phases are present). Note that the volumes of each phase are similar as 2-butanol content increases, which suggests that HLD is not impacted substantially at the weight content of 2-butanol added. Interestingly, the middle phase that consists of a relatively high content of oil and water becomes more transparent as the 2-butanol content increases, which may hint at the stabilizing effect of this alcohol additive. b) 2-butanol helps with stabilization of high content of an oil phase in a WbLS, as shown in these photographs of a WbLS containing 33 wt. % of the oil phase. Note that the white object at the bottom of the vial in each photograph is a stir bar that was not used after initial dissolution and mixing of all components. As shown in the photograph on the left, sedimentation occurs shortly after stirring is completed (ca. one day). This sedimentation appears different than phase separation of oil and water phases and disappears upon gentle mixing (e.g., with a stir bar). Sedimentation was mitigated by adding 2-butanol (second photograph). Upon addition of 2 butanol, the sedimentation ceases and was stable after 7 days (third photograph). The WbLS appears homogeneous and transparent after more than two months after initial preparation (fourth photograph).